\documentclass[aps,prb,showpacs,twocolumn,superscriptaddress]{revtex4-1}
\usepackage{graphicx,amsmath,amssymb}
\usepackage[usenames]{color}
\usepackage{indentfirst}
\usepackage{float}
\usepackage[T1]{fontenc}
\usepackage[colorlinks=true, citecolor=blue, urlcolor=blue, linkcolor=blue ]{hyperref}
\hypersetup{breaklinks=true}
\begin{document}

\bibliographystyle{apsrev4-1}

\title{Exactly Solvable Kondo Lattice Model in Anisotropic Limit}
\author{Wei-Wei Yang}
\affiliation{School of Physical Science and Technology $\&$ Key Laboratory for
Magnetism and Magnetic Materials of the MoE, Lanzhou University, Lanzhou 730000, China}
\author{Jize Zhao}
\affiliation{School of Physical Science and Technology $\&$ Key Laboratory for
Magnetism and Magnetic Materials of the MoE, Lanzhou University, Lanzhou 730000, China}
\author{Hong-Gang Luo}
\email{luohg@lzu.edu.cn}
\affiliation{School of Physical Science and Technology $\&$ Key Laboratory for
Magnetism and Magnetic Materials of the MoE, Lanzhou University, Lanzhou 730000, China}
\affiliation{Beijing Computational Science Research Center, Beijing 100084, China}
\author{Yin Zhong}
\email{zhongy@lzu.edu.cn}
\affiliation{School of Physical Science and Technology $\&$ Key Laboratory for
Magnetism and Magnetic Materials of the MoE, Lanzhou University, Lanzhou 730000, China}
\begin{abstract}
In this paper we introduce an exactly solvable Kondo lattice model without any fine-tuning local gauge symmetry.
This model describes itinerant electrons interplaying with a localized magnetic moment via only longitudinal Kondo exchange.
Its solvability results from conservation of the localized moment at each site, and is valid for arbitrary lattice geometry
and electron filling. A case study on square lattice shows that the ground state is a N\'{e}el antiferromagnetic insulator at half-filling.
At finite temperature, paramagnetic phases including a Mott insulator and correlated metal are found. The former is a melting
antiferromagnetic insulator with a strong short-range magnetic fluctuation, while the latter corresponds to a Fermi liquid-like metal. Monte Carlo simulation and theoretical analysis demonstrate that the transition from paramagnetic phases into the antiferromagnetic insulator is a continuous $2D$ Ising transition. Away from half-filling, patterns of spin stripes (inhomogeneous magnetic order) at weak coupling, and phase separation at strong coupling are predicted. With established Ising antiferromagnetism and spin stripe orders, our model may be relevant to a heavy fermion compound CeCo(In$_{1-x}$Hg$_{x}$)$_{5}$ and novel quantum liquid-crystal order in a hidden order compound URu$_{2}$Si$_{2}$.
\end{abstract}

\maketitle
\section{Introduction}
Exactly solvable quantum many-body models play a fundamental role in understanding exotic quantum states in condensed matter physics\cite{Wen}. In spatial dimension $d>1$, Kitaev's toric-code, honeycomb model and certain $Z_{2}$ lattice gauge field models are prototypical examples\cite{Kitaev1,Kitaev2,Kogut,Prosko}, which shows novel quantum orders like Majorana quantum spin liquid, orthogonal metal, fractionalized Chern insulator, fracton order, Majorana superconductor and many-body localized state\cite{Kitaev2,Nandkishore,Zhong1,Zhong2,Vijay,Parameswaran,Ng,Smith}.
Their solvability is due to the existence of an infinite
number of conserved quantities (for infinite lattice), resulting from intrinsic or emergent local $Z_{2}$ gauge symmetry. However, because of the designed local gauge symmetry, these models are far from the standard ones in condensed matter physics like Hubbard and Kondo lattice models\cite{Hubbard,Tsunetsugu}. Thus, it is highly desirable to find solvable models without any fine-tuning local gauge structure.

In this paper, we introduce an exactly solvable model without any local gauge structure. It is a Kondo lattice model in its anisotropic limit, (also called the Ising-Kondo lattice model)
which describes an itinerant electron interplaying with a localized $f$-electron moment via only longitudinal Kondo exchange \cite{Sikkema}.
\begin{eqnarray}
\hat{H}&&=\sum_{i,j,\sigma}t_{ij}\hat{c}_{i\sigma}^{\dag}\hat{c}_{j\sigma}+\frac{J}{2}\sum_{j\sigma}\hat{S}_{j}^{z}\sigma \hat{c}_{j\sigma}^{\dag}\hat{c}_{j\sigma}\label{eq1}
\end{eqnarray}
$\hat{c}_{j\sigma}$ is the creation operator of conduction electron while $\hat{S}_{j}^{z}$ denotes localized moment of the $f$-electron at site $j$.
$t_{ij}$ is hopping integral between $i,j$ sites and $J$ is the longitudinal Kondo coupling, which is usually antiferromagnetic, i.e. $J>0$. Chemical potential $\mu$ can be added to fix conduction electron's density.

In literature, this model (with $x$-axis anisotropy) is proposed to account for the anomalously small staggered magnetization
and large specific heat jump at hidden order transition in URu$_{2}$Si$_{2}$\cite{Sikkema,Mydosh}. (Due to lack of transverse Kondo coupling, enhancement of effective mass and related Kondo screening are irrelevant.) It can explain
the easy-axis magnetic order and paramagnetic metal or bad metal behaviors in the global phase diagram of heavy fermion compounds\cite{Vojta,Coleman2015,Si,Coleman}.

Importantly, in Eq.~\ref{eq1}, we observe that $f$-electron's spin/localized moment at each site is conservative since $[\hat{S}_{j}^{z},\hat{H}]=0$. Therefore, taking the eigenstates of spin $\hat{S}_{j}^{z}$ as bases, the Hamiltonian is automatically reduced to an effective free fermion model
\begin{eqnarray}
\hat{H}(q)=\sum_{i,j,\sigma}t_{ij}\hat{c}_{i\sigma}^{\dag}\hat{c}_{j\sigma}+\sum_{j\sigma}\frac{J\sigma}{4}q_{j} \hat{c}_{j\sigma}^{\dag}\hat{c}_{j\sigma}\label{eq2}
\end{eqnarray}
with $q$ emphasizing its $q$ dependence and $\hat{S}_{j}^{z}|q_{j}\rangle=\frac{q_{j}}{2}|q_{j}\rangle, q_{j}=\pm1$. Now, the many-body eigenstate of the original model Eq.~\ref{eq1} can be constructed via the single-particle state of effective Hamiltonian Eq.~\ref{eq2} under given configuration of effective Ising spin $\{q_{j}\}$. So, Eq.~\ref{eq1} is exactly solvable and can be considered as an effective spinfull Falicov-Kimball (FK) model\cite{Falicov}. Because the procedure of reduction to the free fermion model involves only local conservation, the above model is solvable for arbitrary lattice geometry, spatial dimension and electron filling, in contrast to isotropic Kondo lattice model\cite{Assaad}, where notorious fermion minus-sign problem prevents exact solution/simulation. Moreover, including the external magnetic field along $z$-axis and Ising term $\sum_{ij}J_{ij}\hat{S}_{i}^{z}\hat{S}_{j}^{z}$ does not change the solvability.

\begin{figure}
\includegraphics[width=1.0\linewidth]{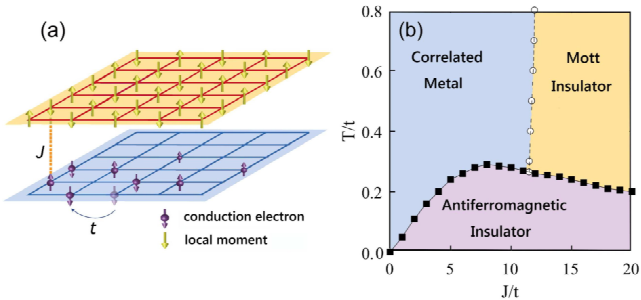}
\caption{\label{fig:1} (a) The Kondo lattice model in an anisotropic limit [Eq.~\ref{eq1}] and (b) its finite temperature phase diagram on a square lattice from LMC. There exist an antiferromagnetic insulator (AI), a Mott insulator (MI) and a correlated metal (CM). The low-$T$ regime of the phase diagram is dominated by AI. At high-$T$, CM in weak coupling is a gapless metallic state while MI in strong coupling exhibits a single-particle excitation gap. The transition from AI to CM or MI is a continuous Ising transition. There only exists a smooth crossover between CM and MI with the opening of a gap at Fermi energy.}
\end{figure}

To illustrate this, we consider the model (\ref{eq1}) on the square lattice with nearest-neighbor-hopping $-t$. (See also Fig.~\ref{fig:1}(a))
In terms of analytical arguments and numerically exact lattice Monte Carlo (LMC) simulation\cite{Czajka}, we have determined its ground state and
its finite temperature phase diagram at half-filling in Fig.~\ref{fig:1}(b). There exist an antiferromagnetic insulator (AI), a Mott insulator (MI) and
a correlated metal (CM). Both AI and MI have gapful single-particle excitations but CM are gapless metallic states. The transition from AI to CM or MI
is a continuous Ising transition, and a smooth crossover appears between CM and MI with the opening of a gap at Fermi energy. The nature of these
phases and transitions will be explored in the main text. For other bipartite lattice like honeycomb lattice, its phase diagram is similar to
Fig.~\ref{fig:1}(b) with the Dirac semimetal replacing CM. When doping away from half-filling, we have established that various patterns
of spin stripes occur at weak coupling and phase separation ultimately dominates at strong coupling. Since Ising antiferromagnetic order has been established in the heavy fermion compound CeCo(In$_{1-x}$Hg$_{x}$)$_{5}$\cite{Stock}, our model could serve as a minimal model for this compound. Moreover, the spin stripe found here may be relevant
to the electronic liquid crystal phenomena in a hidden order compound URu$_{2}$Si$_{2}$\cite{Okazaki}. Finally, our model provides a benchmark for sophisticated numerical techniques like dynamic mean-field theory (DMFT)\cite{Kotliar} and variational cluster approximation (VCA)\cite{Potthoff}.

\section{Ground state}
At zero temperature, when our system is half-filled ($\mu=0$) and on a bipartite lattice (here a square lattice), the ground-state configuration of $q_{j}$ has the two-fold degenerated checkerboard order $q_{j}=\pm(-1)^{j}$(also confirmed by our LMC simulation), according to the theorem proved by Kennedy and Lieb\cite{Kennedy}. (For the generic lattice geometry, the ground-state configuration has to be searched by numerical techniques.) This can be seen as the Ising antiferromagnetic long-ranged order for a localized $f$-electron moment. For conduction electrons, the single-particle Hamiltonian in the ground state is thus
\begin{eqnarray}
\hat{H}&&=-t\sum_{\langle i,j\rangle,\sigma}\hat{c}_{i\sigma}^{\dag}\hat{c}_{j\sigma}+\sum_{j\sigma}(-1)^{j}\frac{J\sigma}{4} \hat{c}_{j\sigma}^{\dag}\hat{c}_{j\sigma}\nonumber\\
&&=\sum_{k\sigma}\left(
                   \begin{array}{cc}
                     \hat{c}_{k\sigma}^{\dag} & \hat{c}_{k+Q,\sigma}^{\dag} \\
                   \end{array}
                 \right)\left(
                          \begin{array}{cc}
                            \varepsilon_{k} & \frac{J\sigma}{4} \\
                            \frac{J\sigma}{4} & \varepsilon_{k+Q} \\
                          \end{array}
                        \right)
                 \left(
                          \begin{array}{c}
                            \hat{c}_{k\sigma} \\
                            \hat{c}_{k+Q,\sigma} \\
                          \end{array}
                        \right)\nonumber
\end{eqnarray}
which is just the familiar antiferromagnetic spin-density-wave (SDW) mean-field Hamiltonian with characteristic wave vector $\vec{Q}=(\pi,\pi)$ in Hubbard-like models\cite{Kusko}. But, we have to emphasize that this one is the exact ground state fermion Hamiltonian for $J>0$. Its quasi-particle dispersion is found to be
$E_{k\sigma}=\pm\sqrt{\varepsilon_{k}^{2}+\frac{J^{2}}{16}}$, which splits the single free conduction electron band into two Hubbard-like bands with direct band gap $\Delta=\frac{J}{2}$. [$\varepsilon_{k}=-2t(\cos k_{x}+\cos k_{y})$ is a free conduction electron dispersion.] We conclude that the ground state of a half-filled model on the square lattice is an insulating antiferromagnetic state with N\'{e}el-like spin order.

\section{Finite-$T$ phase diagram}
\begin{figure}
\includegraphics[width=1.0\linewidth]{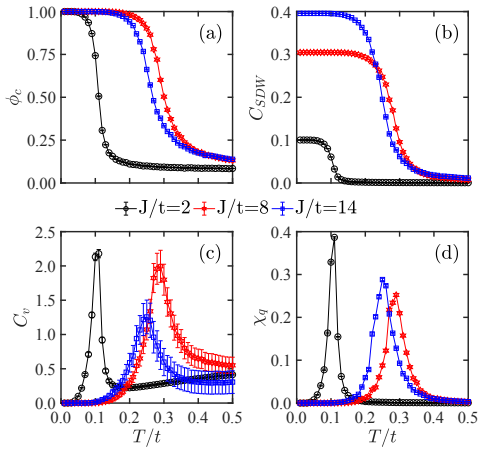}
	\caption{\label{fig:2} (a) The checkerboard order parameter $\phi_{c}$, (b) SDW structure factor $C_{\rm{SDW}}$, (c) specific heat $C_{v}$ and (d) susceptibility $\chi_{q}$ versus temperature $T$ for different Kondo coupling $J$. Both $\phi_{c}$ and $C_{\rm{SDW}}$ saturate at low-$T$, suggesting existence of AI. The vanishing of $C_{\rm{SDW}}$ at high-$T$ signals a transition to paramagnetic phases at critical temperature $T_{c}$. $C_{\mathrm{v}}$ and $\chi_{q}$ also diverge at $T_{c}\simeq0.11t,0.29t,0.25t$ for $J/t=2,8,14$, respectively.}
\end{figure}

At finite temperature, one has to sum over all configurations of the effective Ising spin $\{q_{j}\}$, which can only be performed via Monte Carlo simulation. (See the Appendix for details.) We consider periodic $N_{s}=L\times L$ lattices with $L$ up to $16$. The resulting phase diagram is shown in Fig.~\ref{fig:1}(b). Here, AI is stable at low-$T$ since discrete Ising symmetry is able to break spontaneously into two dimensions $2D$ at finite $T$, and has a thermodynamic transition into CM in weak coupling or MI in strong coupling. The existence of AI is inferred by its checkerboard order parameter $\phi_{c}=\frac{1}{N_{s}}\sum_{j}(-1)^{j}\langle q_{j}\rangle$ and SDW structure factor $C_{\rm{SDW}}=\frac{1}{N_{s}^{2}}\sum_{j,k}(-1)^{j+k}4\langle \hat{s}_{j}^{z}\hat{s}_{k}^{z}\rangle$. ($\hat{s}_{j}^{z}=\frac{1}{2}(\hat{c}_{j\uparrow}^{\dag}\hat{c}_{j\uparrow}-\hat{c}_{j\downarrow}^{\dag}\hat{c}_{j\downarrow})$) In Fig.~\ref{fig:2},
both $\phi_{c}$ and $C_{\rm{SDW}}$ saturate at low-$T$, which suggests the existence of AI. At high-$T$, $C_{\rm{SDW}}$ approaches zero and signals a transition to paramagnetic phases. This agrees with the divergence of specific heat $C_{\mathrm{v}}=\frac{\langle \hat{H}^{2}\rangle-\langle \hat{H}\rangle^{2}}{N_{s}T^{2}}$ and susceptibility $\chi_{q}=\frac{\langle S_{q}^{2}\rangle-\langle S_{q}\rangle^{2}}{T}$ ($S_{q}=\frac{1}{N_{s}^{2}}\sum_{j,k}(-1)^{j+k}q_{i}q_{j}$) at critical temperature $T_{c}$. We note that the maximal $T_{c}$ appears when Kondo coupling is comparable to the band-width of conduction electrons. ($J/t\sim8$) The qualitative physics in AI can be understood by SDW mean-field theory, but such mean-field approximation underestimates thermal fluctuation and leads to unrealistic $T_{c}$.

\subsection{Mott insulator and correlated metal}
\begin{figure}
\flushleft
\includegraphics[width=1.0\linewidth]{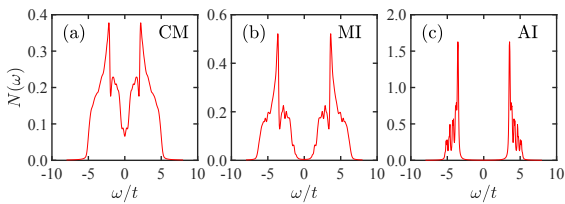}
\caption{\label{fig:3} The density of state for conduction electrons $N(\omega)$ in (a) CM ($J/t=8,T/t=0.4$), (b) MI ($J/t=14,T/t=0.4$) and (c) AI ($J/t=14,T/t=0.1$). }
\end{figure}

\begin{figure}
\includegraphics[width=1.0\linewidth]{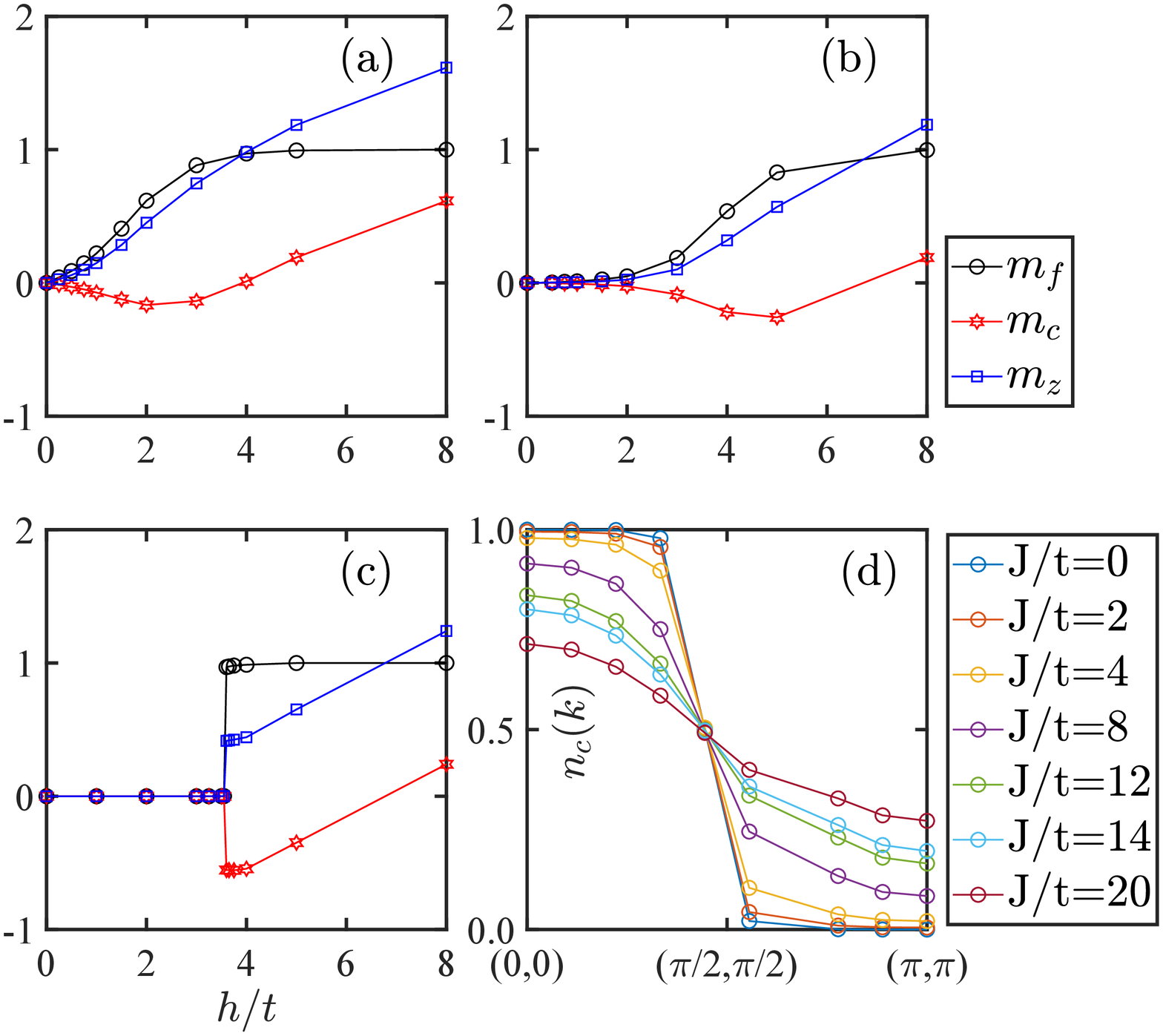}
\caption{\label{fig:7} The magnetization of conduction electrons $m_{c}$, localized $f$-electron moment $m_{f}$ and the total one $m_{z}$ under external magnetic field $h$ for (a) CM ($J/t=8,T/t=0.4$), (b) MI ($J/t=14,T/t=0.4$) and (c) AI ($J/t=14,T/t=0.1$). (d) The conduction electrons' distribution function $n_{c}(k)$ along $(0,0)$ to $(\pi,\pi)$ direction. (T/t=0.4)}
\end{figure}

In the high-$T$ paramagnetic regime, based on the behavior of density of state (DOS) $N(\omega)$ for conduction electrons, magnetization under external magnetic field and conduction electrons' distribution function $n_{c}(k)$, there exist MI and CM. (Fig.~\ref{fig:3} and \ref{fig:7})

MI appears at strong coupling and is a melting AI/SDW without long-ranged order but with (fluctuated) short-ranged order. Such short-ranged order gives rise to a single-particle gap observed in DOS (Fig.~\ref{fig:3}(b)). Intuitively, conduction electrons mediate an antiferromagnetic Ising coupling $\sim \frac{J^{2}}{8t}$ between localized moments. At strong coupling, only temperature itself prevents the formation of long-ranged magnetic order, but short-ranged order survives. Thus, in short-time regime, conduction electrons feel the short-ranged order just as a true long-ranged order, and an excitation gap appears due to the effective molecular field applied to conduction electrons. It is important to note that the short-ranged order has classical Ising feature, thus it does not involve the resonance-valence-bond physics. So, MI here is irrelevant to quantum spin liquids but more like the featureless MI in Bose-Hubbard or FK models\cite{Zhou,Fisher,Antipov}.
(We have checked that MI has very weak $T$-dependence on DOS, contrasting with radical reconstruction around coherent temperature in the usual Kondo insulator\cite{Jarrell}.)

The spectral/single-particle feature of MI can be qualitatively captured by Hubbard-I approximation\cite{Hubbard}, where two Hubbard-like bands are observed and its DOS is similar to LMC's results. In Hubbard-I approximation, the single-electron Green's function is
\begin{equation}
G_{\sigma}(k,\omega)=\frac{1}{\omega-\frac{J^{2}}{16\omega}-\varepsilon_{k}}=\frac{\alpha_{k}^{2}}{\omega-\tilde{E}_{k}^{+}}+\frac{1-\alpha_{k}^{2}}{\omega-\tilde{E}_{k}^{-}}.\nonumber
\end{equation}
Here, the coherent factor $\alpha_{k}^{2}=\frac{1}{2}(1+\frac{\varepsilon_{k}}{\sqrt{\varepsilon_{k}^{2}+J^{2}/4}})$ and the quasiparticle dispersion is $\tilde{E}_{k}^{\pm}=\frac{1}{2}\left[\varepsilon_{k}\pm\sqrt{\varepsilon_{k}+J^{2}/4}\right]$.
Although no antiferromagnetic order exists in MI, the strong local electron correlation (due to Kondo coupling) splits the band and drives the system into correlated MI. Since MI has a single-particle gap, its thermodynamics and transport properties are insulating and show exponential-$T$ behavior. ($C_{\mathrm{v}}$ in MI is extrapolated to vanishing at low-$T$, thus MI are not orthogonal metal-like exotic metals\cite{Nandkishore}.)

At weak coupling, localized $f$-electron moment acts like uncorrelated random potential for conduction electrons, and the average over those disorder potentials leads to correlation correction for the latter one, which is CM. In contrast to gapped MI, CM has finite DOS around zero energy (Fermi energy) without SDW order.(Fig.~\ref{fig:3}(a)) Thus, above $T_{c}$, the gapless CM has linear-$T$ behavior in specific heat ($C_{\mathrm{v}}\sim T$) and it shows Fermi liquid-like behavior. (Fig.~\ref{fig:2}(c))

When the external magnetic field $h$ is applied, in Fig.~\ref{fig:7}, we show the magnetization curvature of conduction electrons ($m_{c}=\sum_{j}\langle\hat{c}_{j\uparrow}^{\dag}\hat{c}_{j\uparrow}-\hat{c}_{j\downarrow}^{\dag}\hat{c}_{j\downarrow}\rangle$), localized $f$-electron moment ($m_{f}=2\sum_{j}\langle S_{j}^{z}\rangle$) and the total one ($m_{z}=m_{c}+m_{f}$). Under large $h$, all states evolve into a fully polarized state. At small $h$, CM shows characteristic metallic linear-$h$ behavior, [Fig.~\ref{fig:7}(a) Pauli-like susceptibility] and both MI and AI have a spin gap. [Figs.~\ref{fig:7}(b) and \ref{fig:7}(c)] A strong first-order transition from AI to a fully polarized state is observed, indicating the absence of a field-induced magnetic quantum phase transition. The quasiparticle behavior in CM and MI can also be inspected in the conduction electrons' distribution function $n_{c}(k)=\langle \hat{c}_{k\sigma}^{\dag}\hat{c}_{k\sigma}\rangle$. [Fig.~\ref{fig:7}(d)] In CM, there is a clear jump in $n_{c}(k)$ around (underlying) Fermi surface [$k=(\pi/2,\pi/2)$] while only smooth evolution exists in MI.

\begin{figure}
\flushleft
\includegraphics[width=1.0\linewidth]{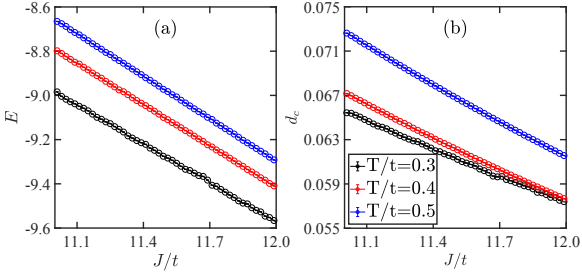}
\caption{\label{fig:4} The energy density $E$ and double occupation $d_{c}$ versus $J$ at different temperature $T$.}
\end{figure}

\begin{figure}
\flushleft
\includegraphics[width=1.0\linewidth]{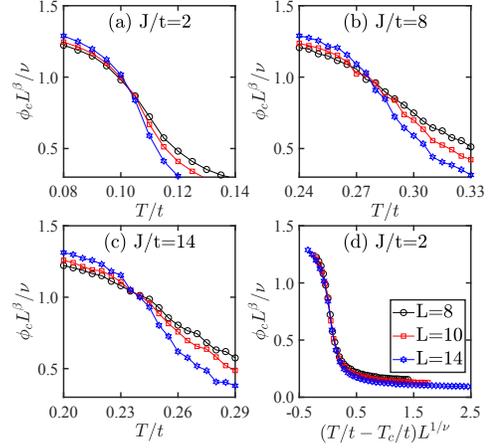}
\caption{\label{fig:6} The finite-size scaling behavior of checkerboard order parameter $\phi_{c}$ for (a) weak coupling ($J/t=2$), (b) intermediate coupling ($J/t=8$), and (c) strong coupling ($J/t=14$). The crossing of the different system size $L=8,10,14$ at $T_{c}$ and the collapse of data in (d) suggests the critical behavior of our model belongs to a $2D$ Ising universality class.}
\end{figure}

There is a smooth crossover between CM and MI as inspecting the evolution of energy density $E=\frac{1}{N_{s}}\langle\hat{H}\rangle$ and double occupation $d_{c}=\frac{1}{N_{s}}\sum_{j}\langle c_{j\uparrow}^{\dag}c_{j\uparrow}c_{j\downarrow}^{\dag}c_{j\downarrow}\rangle$ versus $J$, where $E$ and $d_{c}$ have nearly linear dependence on $J$ and no singularity has been observed(Fig.~\ref{fig:4}).
The transition from MI and CM to AI is continuous as seen from order parameter $\phi_{c}$ and $C_{\mathrm{SDW}}$ around $T_{c}$ [Fig.~\ref{fig:2}(a) and~\ref{fig:2}(b)]. The corresponding histogram for energy density $E$ only shows one peak structure around $T_{c}$ (not shown here), which excludes the strong first-order transition, however the possibility of a very weak discontinuous transition may be still possible. Assuming a continuous phase transition, the finite-size scaling analysis of checkerboard order parameter $\phi_{c}$ in Fig.~\ref{fig:6} suggests that critical behavior of our model belongs to a $2D$ Ising universality class with an order parameter critical exponent $\beta=1/8$ and correlation length critical exponent $\nu=1$. (The standard
scaling form $\phi_{c}L^{\beta/\nu}=g[(T-T_{c})L^{1/\nu}]$ is used to extract $\beta$ and $\nu$\cite{Binder}.)

\subsection{Effective field theory analysis}
We have provided data from LMC simulation, which explores the whole phase diagram and the magnetic transition. In this subsection, we use an effective field theory analysis to understand these phases and related phase transition.

In terms of path integral formalism, we have the following action:
\begin{eqnarray}
S&&=\int d\tau\sum_{\langle i,j\rangle,\sigma}\bar{c}_{i\sigma}(\partial_{\tau}\delta_{ij}-t)c_{j\sigma}\nonumber\\
&&+\int d\tau\left[\frac{J}{2}\sum_{j\sigma}\phi_{j}\sigma\bar{c}_{j\sigma}c_{j\sigma}+i\lambda_{j}(\phi_{j}^{2}-1)\right]\nonumber
\end{eqnarray}
where $\bar{c}_{j\sigma},c_{j\sigma}$ are the Grassmann field for conduction electrons, $\phi_{j}$ denotes the localized moment, and $\lambda_{j}$ is the (dynamic) Lagrangian multiplier. Since above $T_{c}$, the localized moment is disordered, it is reasonable to assume $i\lambda_{j}=m^{2}$ with $m$ being the mass of $\phi_{j}$. This is equivalent to the saddle point approximation for constraint $\phi_{j}^{2}=1$. Then, integrating over $\phi_{j}$, the action only including conduction electrons reads
\begin{eqnarray}
S&&=\int d\tau\sum_{\langle i,j\rangle,\sigma}\bar{c}_{i\sigma}(\partial_{\tau}\delta_{ij}-t)c_{j\sigma}\nonumber\\
&&-\int d\tau\frac{J^{2}}{16m^{2}}\sum_{j\sigma\sigma'}\sigma\bar{c}_{j\sigma}c_{j\sigma}\sigma'\bar{c}_{j\sigma'}c_{j\sigma'}\nonumber
\end{eqnarray}
In the return to Hamiltonian formalism, such action gives rise to a symmetric Hubbard model with effective Hubbard interaction $U_{eff}=\frac{J^{2}}{8m^{2}}$. Therefore, it seems that the high-$T$ paramagnetic states could be related to solutions of half-filled symmetric Hubbard model. However, such relation is not strict due to the saddle point approximation. This explains why at finite-$T$ there is first-order transition between metallic states and MI in Hubbard-like models while in our case, only a smooth crossover appears\cite{Kotliar}. Moreover, when approaching $T_{c}$, the fluctuation effect cannot be neglected, thus it is not reliable to treat our system as an antiferromagnetic Heisenberg model as if $U_{eff}\rightarrow\infty$ for $m^{2}\rightarrow0$. Such feature is consistent with different (spin) symmetry in our model ($Z_{2}$) and Heisenberg model ($SU(2)$).

Alternatively, if we integrate out conduction electrons and set $\phi_{j}=(-1)^{j}\phi_{j}$ to emphasize the dominating antiferromagnetic correlation, the resulting effective theory is the celebrated Hertz-Millis-Moriya theory with $\phi$ being antiferromagnetic Ising order parameter field\cite{Hertz,Millis,Moriya}. Because of the nesting of the Fermi surface on square lattice, antiferromagnetic quantum criticality is avoided and only thermal critical behaviors preserve, (Landau damping due to particle-hole excitation of conduction electrons is subleading and can be neglected away from quantum critical regime) thus the effective theory is replaced by the classic $\phi^{4}$ theory and it corresponds to $2D$ Ising universality class. This agrees with our LMC simulation.

\section{Doping away from half-filling}
\begin{figure}
\includegraphics[width=1.0\linewidth]{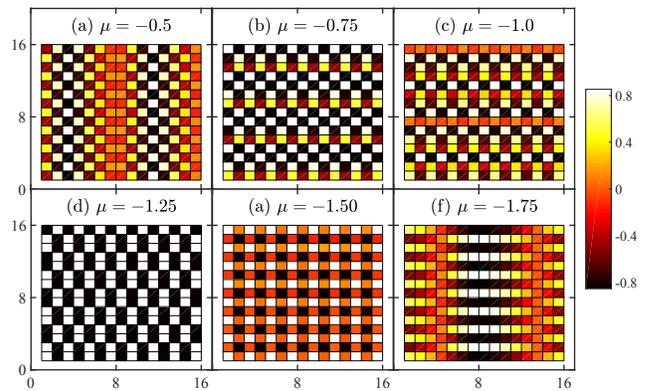}
\caption{\label{fig:8} The spin stripe orders of the local $f$-electron moment ($\langle \hat{S}_{j}^{z}\rangle$) occur in weak coupling regime $J/t=2$ for a different chemical potential $\mu$ ($T/t=0.01$). Here, the value of the color bar denotes the strength of $\langle \hat{S}_{j}^{z}\rangle$.}
\end{figure}

Doping the half-filled system leads to inhomogeneous magnetic order or spin stripe order, as seen in Fig.~\ref{fig:8}. Here, the ground state configuration of the localized $f$-electron moment ($\langle \hat{S}_{j}^{z}\rangle$) versus chemical potential $\mu$ is shown, and various patterns of spin stripes occur to minimize the free energy. Conduction electrons have similar spin stripe structure but with opposite direction and small amplitude. When Kondo coupling $J$ is too large compared with the the band-width of conduction electrons, phase separation ultimately dominates\cite{Czajka2}. These findings are qualitatively consistent with spiral/stripe magnetic phases in the isotropic Kondo lattice model\cite{Costa,Peters}.

\section{Conclusion}
In conclusion, we have provided a solvable lattice fermion Hamiltonian, whose solvability is due to the conserved localized moment at each lattice site. The case study on the square lattice has established the existence of Ising antiferromagnetic order, spin stripe, Fermi liquid-like metallic state and Mott insulator. Importantly, Ising antiferromagnetism and spin stripe orders found here may be relevant to the uniaxial magnetic order in CeCo(In$_{1-x}$Hg$_{x}$)$_{5}$ and the quantum liquid crystal order in URu$_{2}$Si$_{2}$. If frustration is introduced\cite{Czajka}, an exotic metallic phase such as orthogonal metal can be explored\cite{Nandkishore,Hohenadler}. Topological states of matter like topological SDW is also an interesting issue when spin-orbit coupling is added\cite{Zhong2013}. These will be left for our future work.

\section*{Acknowledgments}
We thank Hantao Lu, Yuehua Su and Yu Liu for helpful discussion on the phase diagram and nature of each phase. This research was supported in part by NSFC under Grant No.~$11704166$, No.~$11834005$, No.~$11874188$ and the Fundamental Research Funds for the Central Universities.

Y.Z. proposed the project, Y.Z. and W.Y. carried out the calculation.

\appendix
\section{Details of Lattice Monte Carlo simulation}
When exploring the finite temperature properties, one has to sum over all configurations of effective Ising spin $\{q_{j}\}$, which can only be performed via Monte Carlo simulation\cite{Czajka}.
\subsection{ Monte Carlo simulation}
Firstly, we know that the equilibrium state thermodynamics is controlled by the partition function
\begin{equation}
\mathcal{Z}=\mathrm{Tr}e^{-\beta \hat{H}}=\mathrm{Tr}_{c}\mathrm{Tr}_{S}e^{-\beta \hat{H}}=\sum_{\{q_{j}\}}\mathrm{Tr}_{c}e^{-\beta \hat{H}(q)}.\nonumber
\end{equation}

Here, the trace is split into $c$-fermion and $\hat{S}^{z}$, where the latter is transformed into the summation over all possible configuration $\{q_{j}\}$. For each single-particle Hamiltonian $\hat{H}(q)$, it can be easily diagonalized into
\begin{equation}
\hat{H}(q)=\sum_{n}E_{n}\hat{d}_{n}^{\dag}\hat{d}_{n}\nonumber
\end{equation}
where $E_{n}$ is the single-particle energy level and $\hat{d}_{n}$ is the quasi-particle. The fermion $\hat{d}_{n}$ is related into $\hat{c}_{j}$
via
\begin{equation}
\hat{c}_{j}=|0\rangle\langle j|=\sum_{n}|0\rangle\langle j|n\rangle\langle n|=\sum_{n}|0\rangle\langle n|\langle j|n\rangle=\sum_{n}\hat{d}_{n}\phi_{n}^{j}\nonumber.
\end{equation}
with $\phi_{n}^{j}\equiv\langle j|n\rangle$ while the creation operator $\hat{c}_{j}^{\dag}$ has
\begin{equation}
\hat{c}_{j}^{\dag}=|j\rangle \langle 0|=\sum_{n}|n\rangle\langle n|j\rangle\langle0|=\sum_{n}|n\rangle\langle 0|\langle n|j\rangle=\sum_{n}\hat{d}_{n}^{\dag}(\phi_{n}^{j})^{\ast}\nonumber.
\end{equation}

Now, the trace over $c$-fermion can be obtained as
\begin{eqnarray}
\mathrm{Tr}_{c}e^{-\beta \hat{H}(q)}&&=\sum_{n}\langle n|e^{-\beta \sum_{m}E_{m}\hat{d}_{m}^{\dag}\hat{d}_{m}}|n\rangle\nonumber\\
&&=\prod_{m}\sum_{n}\langle n|e^{-\beta E_{m}\hat{d}_{m}^{\dag}\hat{d}_{m}}|n\rangle\nonumber\\
&&=\prod_{m}\sum_{n}\sum_{n_{m}=0,1}\delta_{n,m}(1+e^{-\beta E_{m}n_{m}})\nonumber\\
&&=\prod_{m}\sum_{n}\delta_{n,m}(1+e^{-\beta E_{m}})\nonumber\\
&&=\prod_{m}(1+e^{-\beta E_{m}}).\nonumber
\end{eqnarray}
This is the familiar result for free fermion, however one should keep in mind that $E_{n}$ actually depends on the effective Ising spin configuration $\{q_{j}\}$, thus we write $E_{n}(q)$ to emphasize this fact.

So, the partition function reads
\begin{equation}
\mathcal{Z}=\sum_{\{q_{j}\}}\prod_{n}(1+e^{-\beta E_{n}(q)})=\sum_{\{q_{j}\}}e^{-\beta F(q)}\nonumber
\end{equation}
where we have defined an effective free energy
\begin{equation}
F(q)=-T\sum_{n}\ln(1+e^{-\beta E_{n}(q)}).\label{eq13}
\end{equation}
In this situation, we can explain $e^{-\beta F(q)}$ or $\rho(q)=\frac{1}{\mathcal{Z}}e^{-\beta F(q)}$ as an effective Boltzmann weight for each configuration of $\{q_{j}\}$ and this can be used to perform Monte Carlo simulation just like the classic Ising model. Specifically, we start with random or chosen configuration of $\{q_{j}\}$, then try to flip each $q_{j}$ to $-q_{j}$. The relative probability for such flip is determined by effective Boltzmann weight
\begin{equation}
r=\frac{e^{-\beta F_{new}(q)}}{e^{-\beta F_{old}(q)}}=e^{-\beta (F_{new}(q)-F_{old}(q))}\label{eq14}
\end{equation}
where $F_{old}(q),F_{new}(q)$ denotes the effective free energy before and after flip. This is the weight used in the classic Metropolic importance sampling. Alternatively, one can use the so-called bath algorithm, which means
\begin{equation}
\tilde{r}=\frac{r}{1+r}\label{eq15}
\end{equation}
such that the probability is seemingly normalized. Then, we generate a random number $a$ from uniform distribution $[0,1]$ and compare this with $\tilde{r}$. If $\tilde{r}>a$, the flip of $q_{j}$ is accepted and $F_{old}$ is updated into $F_{new}$, otherwise we reset $q_{j}$ to its original value before the flip.
By trying to flip $q_{j}$ over all sites, this is called a sweep and doing such sweep many times, the system can be equilibrium and the next sweeps are used to calculate physical observable.
\subsection{On physical quantities}
To calculate physical quantities, we consider generic operator $\hat{O}$, which can be split into part with only Ising spin $\{q_{j}\}$ and another part with fermions,
\begin{equation}
\hat{O}=\hat{O}^{c}+\hat{O}^{q}.\nonumber
\end{equation}
Then, its expectation value in the equilibrium ensemble reads
\begin{equation}
\langle\hat{O}\rangle=\langle\hat{O}^{c}\rangle+\langle\hat{O}^{q}\rangle=\frac{\mathrm{Tr} \hat{O}^{c}e^{-\beta \hat{H}}}{\mathrm{Tr} e^{-\beta \hat{H}}}+\frac{\mathrm{Tr} \hat{O}^{q}e^{-\beta \hat{H}}}{\mathrm{Tr} e^{-\beta \hat{H}}}\nonumber
\end{equation}
For $\hat{O}^{q}$, we have
\begin{eqnarray}
\langle\hat{O}^{q}\rangle&&=\frac{\sum_{\{q_{j}\}}\hat{O}^{q}(q)e^{\beta h\sum_{j}q_{j}}\mathrm{Tr}_{c}e^{-\beta \hat{H}(q)}}{\sum_{\{q_{j}\}}e^{-\beta F(q)}}\nonumber\\
&&=\frac{\sum_{\{q_{j}\}}\hat{O}^{q}(q)e^{-\beta F(q)}}{\sum_{\{q_{j}\}}e^{-\beta F(q)}}\nonumber\\
&&=\sum_{\{q_{j}\}}\hat{O}^{q}(q)\rho(q)\nonumber.
\end{eqnarray}
In the Metropolic importance sampling algorithm, the above equation means we can use the simple average to estimate the expectation value like
\begin{equation}
\langle\hat{O}^{q}\rangle\simeq \frac{1}{N_{m}}\sum_{\{q_{j}\}} \hat{O}^{q}(q)\label{eq16}
\end{equation}
where $N_{m}$ is the number of sampling and the sum is over each configuration. $\hat{O}^{q}(q)$ is a number since we always work on the basis of $\{q_{j}\}$.

For $\hat{O}^{c}$,
\begin{equation}
\langle\hat{O}^{c}\rangle=\frac{\sum_{\{q_{j}\}}e^{\beta h\sum_{j}q_{j}}\mathrm{Tr}_{c}\hat{O}^{c}(q)e^{-\beta \hat{H}(q)}}{\sum_{\{q_{j}\}}e^{-\beta F(q)}}\nonumber
\end{equation}
and we can insert $\frac{e^{-\beta F(q)}}{e^{-\beta F(q)}}$ in the numerator, which leads to
\begin{eqnarray}
\langle\hat{O}^{c}\rangle&&=\sum_{\{q_{j}\}}\frac{\mathrm{Tr}_{c}\hat{O}^{c}(q)e^{-\beta \hat{H}(q)}}{\mathrm{Tr}_{c}e^{-\beta \hat{H}(q)}}\frac{e^{-\beta F(q)}}{\sum_{\{q_{j}\}}e^{-\beta F(q)}}\nonumber\\
&&=\sum_{\{q_{j}\}}\frac{\mathrm{Tr}_{c}\hat{O}^{c}(q)e^{-\beta \hat{H}(q)}}{\mathrm{Tr}_{c}e^{-\beta \hat{H}(q)}}\rho(q)\nonumber\\
&&=\sum_{\{q_{j}\}}\langle\langle\hat{O}^{c}(q)\rangle\rangle\rho(q)\nonumber.
\end{eqnarray}
This means
\begin{equation}
\langle\hat{O}^{c}\rangle\simeq \frac{1}{N_{m}}\sum_{\{q_{j}\}} \langle\langle\hat{O}^{c}(q)\rangle\rangle,\label{eq17}
\end{equation}
where $\langle\langle\hat{O}^{c}(q)\rangle\rangle=\frac{\mathrm{Tr}_{c}\hat{O}^{c}(q)e^{-\beta \hat{H}(q)}}{\mathrm{Tr}_{c}e^{-\beta \hat{H}(q)}}$ is calculated based on the Hamiltonian $\hat{H}(q)$. More practically, such statement means if fermions are involved, one can just calculate with $\hat{H}(q)$. Then, average over all sampled configuration gives rise to the wanted results.

\subsection{Energy, correlation and others}
The total energy of our system is an essential quantity and can be calculated as
\begin{eqnarray}
\langle\hat{H}\rangle&&\simeq \frac{1}{N_{m}}\sum_{\{q_{j}\}} \left[\langle\langle\hat{H}(q)\rangle\rangle\right]\nonumber\\
&&=\frac{1}{N_{m}}\sum_{\{q_{j}\}}\left[\sum_{n}E_{n}(q)f_{F}(E_{n}(q))\right].\nonumber
\end{eqnarray}
Here, the $\langle\langle\hat{H}(q)\rangle\rangle$ is just the summation over all quasi-particles for given configuration. Now, the specific heat $C_{\mathrm{v}}$ can be found by $C_{\mathrm{v}}=\frac{\langle\hat{H}^{2}\rangle-\langle\hat{H}\rangle^{2}}{T^{2}}$, which means
\begin{eqnarray}
C_{\mathrm{v}}&&\simeq \frac{\frac{1}{N_{m}}\sum_{\{q_{j}\}}\langle\langle \hat{H}^{2}(q)\rangle\rangle}{T^{2}}-\frac{\langle \hat{H}\rangle^{2}}{T^{2}}\nonumber
\end{eqnarray}
where
\begin{eqnarray}
\langle\langle \hat{H}^{2}(q)\rangle\rangle&&=\left(\sum_{n}E_{n}(q)f_{F}(E_{n}(q))\right)^{2}\nonumber\\
&&+\sum_{n}E_{n}^{2}(q)f_{F}(E_{n}(q))\left[1-f_{F}(E_{n}(q))\right]\nonumber
\end{eqnarray}

Next, we consider the spin-spin correlation function $S_{qq}(Q)$, which is defined as
\begin{equation}
S_{qq}(Q)=\frac{1}{N_{s}^{2}}\sum_{ij}e^{iQ(R_{i}-R_{j})}\langle q_{i}q_{j}\rangle.\nonumber
\end{equation}
This object is designed to detect the static or dynamic (fluctuation) order with the characteristic wave-vector $Q$. If spin orders in certain $Q$, the value of the corresponding $S_{qq}(Q)$ should reach $\mathcal{O}(1)$.
Now, using Eq.~\ref{eq16}, we obtain
\begin{equation}
S_{qq}(Q)\simeq\frac{1}{N_{m}}\sum_{\{q_{j}\}}\left[\frac{1}{N_{s}^{2}}\sum_{ij}e^{iQ(R_{i}-R_{j})}q_{q}q_{j}\right].\nonumber
\end{equation}
Usually, we may also use its susceptibility
\begin{eqnarray}
\chi_{qq}(Q)&&=\frac{\langle S_{qq}^{2}(Q)\rangle-\langle S_{qq}(Q)\rangle^{2}}{T}\nonumber\\
&&\simeq\frac{1}{T}\frac{1}{N_{m}}\sum_{\{q_{j}\}}\left[\frac{1}{N_{s}^{2}}\sum_{ij}e^{iQ(R_{i}-R_{j})}q_{q}q_{j}\right]^{2}\nonumber\\
&&-\frac{1}{T}\left(\frac{1}{N_{m}}\sum_{\{q_{j}\}}\left[\frac{1}{N_{s}^{2}}\sum_{ij}e^{iQ(R_{i}-R_{j})}q_{q}q_{j}\right]\right)^{2}\nonumber
\end{eqnarray}
to locate the position of long-ranged order. Obviously, if orders appear, $\chi_{qq}(Q)$ has to diverge for certain $Q$ at some critical temperature $T_{c}$. At $T_{c}$, the specific heat $C_{\mathrm{v}}$ diverges as well.

For total $c$-fermion density, in terms of Eq.~\ref{eq17}, it is easy to show that
\begin{equation}
n_{c}=\frac{1}{N_{s}}\sum_{j}\langle \hat{c}_{j}^{\dag}\hat{c}_{j}\rangle\simeq\frac{1}{N_{m}}\sum_{\{q_{j}\}}\left[\frac{1}{N_{s}}\sum_{n}f_{F}(E_{n}(q))\right]\nonumber.
\end{equation}
Since we work on grand canonical ensemble, if chemical potential is setting to zero, the model is symmetric and $c$-fermion should be half-filled. So, in this situation, $n_{c}$ is $1$, irrespective of temperature.

If we are interested in the density of state $N(\omega)$ of $c$-fermion, we know that for $\hat{H}(q)$, it is given by
\begin{equation}
N(\omega,q)=\frac{1}{N_{s}}\sum_{n}\delta(\omega-E_{n}(q))\nonumber,
\end{equation}
thus,
\begin{equation}
N(\omega)\simeq\frac{1}{N_{m}N_{s}}\sum_{\{q_{j}\}}\sum_{n}\delta(\omega-E_{n}(q))\nonumber.
\end{equation}

Finally, when we calculate fermion's correlation function like $\langle \hat{c}_{i}^{\dag}\hat{c}_{j}\hat{c}_{k}^{\dag}\hat{c}_{l}\rangle$,
\begin{equation}
\langle \hat{c}_{i}^{\dag}\hat{c}_{j}\hat{c}_{k}^{\dag}\hat{c}_{l}\rangle=\frac{1}{N_{m}}\sum_{\{q_{j}\}}\langle\langle \hat{c}_{i}^{\dag}\hat{c}_{j}\hat{c}_{k}^{\dag}\hat{c}_{l}\rangle\rangle.\label{eq8}
\end{equation}
Then, using the Wick theorem for these free fermions, we get
\begin{eqnarray}
\langle\langle\hat{c}_{i}^{\dag}\hat{c}_{j}\hat{c}_{k}^{\dag}\hat{c}_{l}\rangle\rangle&&=\langle\langle\hat{c}_{i}^{\dag}\hat{c}_{j}\rangle\rangle\langle\langle\hat{c}_{k}^{\dag}\hat{c}_{l}\rangle\rangle+\langle\langle\hat{c}_{i}^{\dag}\hat{c}_{l}\rangle\rangle\langle\langle\hat{c}_{j}\hat{c}_{k}^{\dag}\rangle\rangle.\label{eq9}
\end{eqnarray}
Next, for each one-body correlation function like $\langle\langle\hat{c}_{i}^{\dag}\hat{c}_{j}\rangle\rangle$, one can transform these objects into their quasiparticle basis,
\begin{eqnarray}
g_{ij}^{q}\equiv\langle\langle \hat{c}_{i}^{\dag}\hat{c}_{j}\rangle\rangle&&=\sum_{m,n}\langle\langle\hat{d}_{m}^{\dag}\hat{d}_{n}\rangle\rangle(\phi_{m}^{i})^{\ast}\phi_{n}^{j}\nonumber\\
&&=\sum_{m,n}f_{F}(E_{n}(q))\delta_{m,n}(\phi_{m}^{i})^{\ast}\phi_{n}^{j}\nonumber\\
&&=\sum_{n}f_{F}(E_{n}(q))(\phi_{n}^{i})^{\ast}\phi_{n}^{j}\nonumber,
\end{eqnarray}
Similarly, we have
\begin{equation}
\langle\langle \hat{c}_{i}\hat{c}_{j}^{\dag}\rangle\rangle=\sum_{n}(1-f_{F}(E_{n}(q)))\phi_{n}^{i}(\phi_{n}^{j})^{\ast}=\delta_{ij}-g_{ji}^{q}.\nonumber
\end{equation}
Then, combining Eq.~\ref{eq9} and inserting this expression into the Eq.~\ref{eq8}, we can calculate the fermion's correlation function.

For instance, the spin-density-wave structure factor is calculated as
\begin{eqnarray}
C_{\mathrm{SDW}}&&=\frac{1}{N_{m}}\sum_{\{q_{j}\}}\frac{1}{4N_{s}^{2}}\sum_{j,k,\sigma,\sigma'}(-1)^{j+k}\langle\langle(\sigma\hat{c}_{j\sigma}^{\dag}\hat{c}_{j\sigma})(\sigma'\hat{c}_{k\sigma'}^{\dag}\hat{c}_{k\sigma'})\rangle\rangle\nonumber\\
&&=\frac{1}{N_{m}}\sum_{\{q_{j}\}}\frac{1}{4N_{s}^{2}}\sum_{j,k}(-1)^{j+k}\times[g_{jj,\uparrow}^{q}g_{kk,\uparrow}^{q}\nonumber\\
&&+g_{jk,\uparrow}^{q}(\delta_{jk}-g_{kj,\uparrow}^{q})-q_{jj,\uparrow}^{q}g_{kk,\downarrow}^{q}-q_{jj,\downarrow}^{q}g_{kk,\uparrow}^{q}\nonumber\\
&&+g_{jj,\downarrow}^{q}g_{kk,\downarrow}^{q}+g_{jk,\downarrow}^{q}(\delta_{jk}-g_{kj,\downarrow}^{q})]
\end{eqnarray}
\section{Effect of external magnetic field: Magnetization}
If $z$-axis external magnetic field is included, our model is changed to
\begin{eqnarray}
\hat{H}=&&-t\sum_{\langle i,j\rangle,\sigma}\hat{c}_{i\sigma}^{\dag}\hat{c}_{j\sigma}+\frac{J}{2}\sum_{j\sigma}S_{j}^{z}\sigma \hat{c}_{j\sigma}^{\dag}\hat{c}_{j\sigma}\nonumber\\
&&-h\sum_{j}\left(S_{j}^{z}+\frac{1}{2}\sum_{\sigma}\sigma \hat{c}_{j\sigma}^{\dag}\hat{c}_{j\sigma}\right)
\end{eqnarray}
With eigenstates of spin $S_{j}^{z}$ as bases, the corresponding effective free fermion model reads
\begin{eqnarray}
\hat{H}=-t\sum_{\langle i,j\rangle,\sigma}\hat{c}_{i\sigma}^{\dag}\hat{c}_{j\sigma}+\sum_{j\sigma}\left(\frac{J}{4}q_{j}-\frac{h}{2}\right)\sigma \hat{c}_{j\sigma}^{\dag}\hat{c}_{j\sigma},
\end{eqnarray}
which is readily to be simulated by LMC with effective free energy $F(q)=-T\sum_{n}\ln(1+e^{-\beta E_{n}(q)})-h/2\sum_{j}q_{j}$. In the main text, by using above formalism, we have calculated the magnetization of conduction electron and localized moment under external field $h$.

\section{Phase separation}
\begin{figure}
\includegraphics[width=1.2\linewidth]{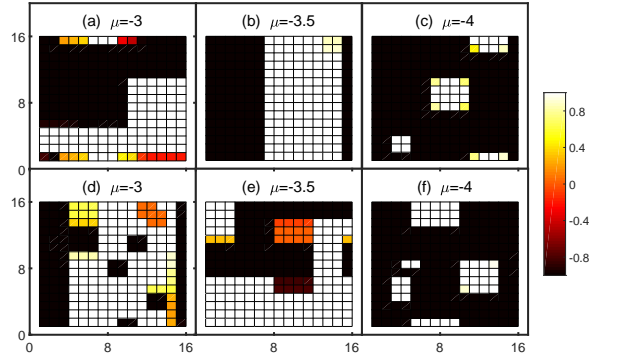}
\caption{\label{fig:9} The phase separation at intermediate ((a), (b) and (c) with $J/t=8$) and strong coupling ((d), (e) and (f) with $J/t=14$) situation.}
\end{figure}

When doping away from half-filling, the microscopic phase separation is a leading effect for intermediate and large Kondo coupling. An example is given in Fig.~\ref{fig:9}, where the phase separation is clearly seen with different ferromagnetic droplets.

In addition, we note that if conduction electron's density is very small ($n_{c}\sim0.1$), ferromagnetism is stable both in weak and strong coupling regime.
\section{Mean-field approximation for SDW state}
The SDW mean-field Hamiltonian can be obtained by decoupling the Kondo coupling term via
\begin{equation}
\sum_{\sigma}S_{j}^{z}\sigma \hat{c}_{j\sigma}^{\dag}\hat{c}_{j\sigma}\simeq (-1)^{j}\frac{m_{f}}{2}\sigma \hat{c}_{j\sigma}^{\dag}\hat{c}_{j\sigma}+(-1)^{j+1}m_{c}S_{j}^{z}+\frac{m_{f}m_{c}}{2}\nonumber
\end{equation}
where the magnetic order parameters are defined by
\begin{eqnarray}
&&\langle S_{j}^{z}\rangle=(-1)^{j}\frac{m_{f}}{2}\nonumber\\
&&\sum_{\sigma}\sigma\langle \hat{c}_{j\sigma}^{\dag}\hat{c}_{j\sigma}\rangle=(-1)^{j+1}m_{c},
\end{eqnarray}
thus, the mean-field Hamiltonian reads
\begin{eqnarray}
\hat{H}&&=-t\sum_{\langle i,j\rangle,\sigma}\hat{c}_{i\sigma}^{\dag}\hat{c}_{j\sigma}+\frac{Jm_{f}}{4}\sum_{j\sigma}(-1)^{j}\sigma \hat{c}_{j\sigma}^{\dag}\hat{c}_{j\sigma}\nonumber\\
&&+\frac{J}{2}(-1)^{j+1}m_{c}\sum_{j}S_{j}^{z}+\frac{J}{2}\sum_{j}\frac{m_{f}m_{c}}{2}\nonumber\\
&&=\sum_{k\sigma}\varepsilon_{k}\hat{c}_{k\sigma}^{\dag}\hat{c}_{k\sigma}+\frac{Jm_{f}}{4}\sum_{k\sigma}\hat{c}_{k+Q\sigma}^{\dag}\hat{c}_{k\sigma}\nonumber\\
&&+\frac{J}{2}m_{c}\sum_{j}(-1)^{j+1}S_{j}^{z}+\frac{J}{2}\sum_{j}\frac{m_{f}m_{c}}{2}\nonumber\\
&&=\sum_{k\sigma}'\left(
                   \begin{array}{cc}
                     \hat{c}_{k\sigma}^{\dag} & \hat{c}_{k+Q,\sigma}^{\dag} \\
                   \end{array}
                 \right)\left(
                          \begin{array}{cc}
                            \varepsilon_{k} & \frac{Jm_{f}\sigma}{4} \\
                            \frac{Jm_{f}\sigma}{4} & \varepsilon_{k+Q} \\
                          \end{array}
                        \right)
                 \left(
                          \begin{array}{c}
                            \hat{c}_{k\sigma} \\
                            \hat{c}_{k+Q,\sigma} \\
                          \end{array}
                        \right)\nonumber\\
&&+\frac{J}{2}m_{c}\sum_{j}(-1)^{j+1}S_{j}^{z}+\frac{J}{2}\sum_{j}\frac{m_{f}m_{c}}{2}\nonumber.
\end{eqnarray}
Here, the summation over momentum $k$ is restricted into the folded first Brillouin zone (for square lattice, the area is interior surrounded by  $k_{y}=-k_{x}\pm\pi,k_{y}=k_{x}\pm\pi$) Therefore, its free energy can be found as
\begin{eqnarray}
F=&&-T\sum_{k\sigma}'\left[\ln(1+e^{-\beta E_{k+}})+\ln(1+e^{-\beta E_{k-}})\right]+N_{s}\frac{Jm_{f}m_{c}}{4}\nonumber\\
&&-TN_{s}\ln\left(2\cosh\frac{\beta Jm_{c}}{4}\right)
\end{eqnarray}
with $E_{k\pm}=\pm\sqrt{\varepsilon_{k}^{2}+\frac{J^{2}m_{f}^{2}}{16}}$. So, the mean-field self-consistent equations can be derived by
$\frac{\partial F}{\partial m_{c}}=0,\frac{\partial F}{\partial m_{f}}=0$,
\begin{equation}
m_{f}=\tanh\frac{Jm_{c}}{4T}
\end{equation}
\begin{equation}
m_{c}=\frac{Jm_{f}}{4}\frac{2}{N_{s}}\sum_{k}'\frac{\tanh\frac{E_{k+}}{2T}}{E_{k+}}.
\end{equation}
From these two equations, one can obtain the equation for SDW critical temperature $T_{c}^{mf}$ in the mean-field approximation.
\begin{figure}
\flushleft
\includegraphics[width=1.0\linewidth]{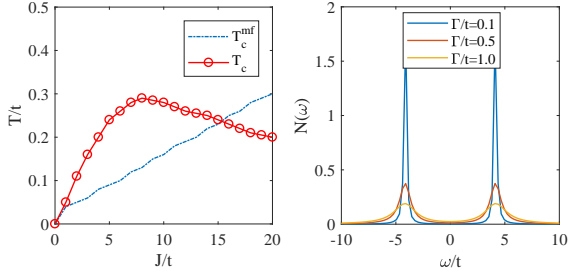}
\caption{\label{fig:5} (a) The mean-field critical temperature $J_{c}^{mf}$ versus $J$. For comparison, exact critical temperature $T_{c}$ from LMC is also shown. (b) Density of state $N(\omega)$ of conduction electron in MI with Hubbard-I approximation.}
\end{figure}
In Fig.~\ref{fig:5}(a), we show the SDW critical temperature $T_{c}^{mf}$ versus $J$, which has linear dependence on $J$. However, since no thermal fluctuation is included, the critical temperature $T_{c}^{mf}$ shows a wrong tendency compared to the exact critical temperature $T_{c}$ from LMC.

\section{Hubbard-I approximation for MI}
In the paramagnetic MI, one can follow the Hubbard-I approximation to give a rough solution for MI itself\cite{Hubbard}. To this purpose, we use equation of motion (EOM) formalism and define the retarded Green's function as
\begin{equation}
G_{ij,\sigma}(\omega)=\langle\langle \hat{c}_{i\sigma}|\hat{c}_{j\sigma}^{\dag}\rangle\rangle_{\omega}.\nonumber
\end{equation}
By using the standard EOM relation,
\begin{equation}
\omega\langle\langle \hat{A}|\hat{B}\rangle\rangle_{\omega}=\langle [\hat{A},\hat{B}]_{+}\rangle+\langle\langle[\hat{A},\hat{H}]_{-}|\hat{B}\rangle\rangle_{\omega}\nonumber
\end{equation}
for our model, it follows that
\begin{equation}
\omega\langle\langle \hat{c}_{i\sigma}|\hat{c}_{j\sigma}^{\dag}\rangle\rangle_{\omega}=\delta_{ij}-t\Delta_{im}\langle\langle \hat{c}_{m\sigma}|\hat{c}_{j\sigma}^{\dag}\rangle\rangle_{\omega}+\frac{J}{2}\sigma\langle\langle \hat{S}_{i}^{z}\hat{c}_{i\sigma}|\hat{c}_{j\sigma}^{\dag}\rangle\rangle_{\omega}.\nonumber
\end{equation}
Here, $\Delta_{im}$ denotes $m$ is the nearest-neighbor site of $i$. For $\langle\langle \hat{S}_{i}^{z}\hat{c}_{i\sigma}|\hat{c}_{j\sigma}^{\dag}\rangle\rangle_{\omega}$, we have
\begin{eqnarray}
\omega\langle\langle \hat{S}_{i}^{z}\hat{c}_{i\sigma}|\hat{c}_{j\sigma}^{\dag}\rangle\rangle_{\omega}&&=\langle \hat{S}_{i}^{z}\rangle\delta_{ij}-t\Delta_{il}\langle\langle \hat{S}_{i}^{z}\hat{c}_{l\sigma}|\hat{c}_{j\sigma}^{\dag}\rangle\rangle_{\omega}\nonumber\\
&&+\frac{J}{8}\sigma\langle\langle \hat{c}_{i\sigma}|\hat{c}_{j\sigma}^{\dag}\rangle\rangle_{\omega}.\nonumber
\end{eqnarray}
If no further EOM are involved, to close EOM, we have to decouple $\langle\langle \hat{S}_{i}^{z}\hat{c}_{l\sigma}|\hat{c}_{j\sigma}^{\dag}\rangle\rangle_{\omega}$ as
\begin{equation}
\langle\langle \hat{S}_{i}^{z}\hat{c}_{l\sigma}|\hat{c}_{j\sigma}^{\dag}\rangle\rangle_{\omega}\simeq\langle \hat{S}_{i}^{z}\rangle\langle\langle\hat{c}_{l\sigma}|\hat{c}_{j\sigma}^{\dag}\rangle\rangle_{\omega}
\end{equation}
In paramagnetic strong coupling regime, there is no magnetic order, thus $\langle \hat{S}_{i}^{z}\rangle=0$. Meanwhile, the contribution from $\langle\langle \hat{S}_{i}^{z}\hat{c}_{l\sigma}|\hat{c}_{j\sigma}^{\dag}\rangle\rangle_{\omega}$ vanishes due to the decoupling and the above equation has a complete solution:
\begin{equation}
\left(\omega-\frac{J^{2}}{16\omega}\right)\langle\langle \hat{c}_{i\sigma}|\hat{c}_{j\sigma}^{\dag}\rangle\rangle_{\omega}=\delta_{ij}-t\Delta_{im}\langle\langle \hat{c}_{m\sigma}|\hat{c}_{j\sigma}^{\dag}\rangle\rangle_{\omega},\nonumber
\end{equation}
which can be written as
\begin{equation}
\left(\omega-\frac{J^{2}}{16\omega}\right)G_{ij,\sigma}(\omega)=\delta_{ij}-t\Delta_{im}G_{mj,\sigma}(\omega)
\end{equation}
Now, performing the Fourier transformation
\begin{equation}
G_{ij,\sigma}(\omega)=\frac{1}{N_{s}}\sum_{k}e^{ik(R_{i}-R_{j})}G_{\sigma}(k,\omega),\nonumber
\end{equation}
we have
\begin{eqnarray}
&&\sum_{k}\left(\omega-\frac{J^{2}}{16\omega}\right)G_{\sigma}(k,\omega)e^{ik(R_{i}-R_{j})}=\sum_{k}e^{ik(R_{i}-R_{j})}\nonumber\\
&&-t\sum_{k}\Delta_{im}G_{\sigma}(k,\omega)e^{ik(R_{m}-R_{j}+R_{i}-R_{i})}.
\end{eqnarray}
Here,$-t\Delta_{im}e^{ik(R_{m}-R_{i})}=-t\sum_{\delta}e^{-ik\delta}=\varepsilon_{k}$ and we have
\begin{eqnarray}
&&\sum_{k}\left[\left(\omega-\frac{J^{2}}{16\omega}-\varepsilon_{k}\right)G_{\sigma}(k,\omega)-1\right]e^{ik(R_{i}-R_{j})}=0.\nonumber
\end{eqnarray}
Since $R_{i},R_{j}$ is arbitrary, the above equation is valid if $\left(\omega-\frac{J^{2}}{16\omega}-\varepsilon_{k}\right)G_{\sigma}(k,\omega)-1=0$, which gives the single-particle Green's function as
\begin{equation}
G_{\sigma}(k,\omega)=\frac{1}{\omega-\frac{J^{2}}{16\omega}-\varepsilon_{k}}=\frac{\alpha_{k}^{2}}{\omega-\tilde{E}_{k}^{+}}+\frac{1-\alpha_{k}^{2}}{\omega-\tilde{E}_{k}^{-}}.\label{eq_h1}
\end{equation}
Here, the coherent factor $\alpha_{k}^{2}=\frac{1}{2}\left(1+\frac{\varepsilon_{k}}{\sqrt{\varepsilon_{k}^{2}+J^{2}/4}}\right)$ and
\begin{equation}
\tilde{E}_{k}^{\pm}=\frac{1}{2}\left[\varepsilon_{k}\pm\sqrt{\varepsilon_{k}+J^{2}/4}\right].
\end{equation}
Therefore, although no SDW state appears, the band of conduction electron has been split into two Hubbard-like bands. This is driven by the local Kondo interaction and the resulting state is MI if half-filling is satisfied. Moreover, the DOS in the present Hubbard-I approximation can be obtained by
\begin{equation}
N(\omega)=\frac{1}{N_{s}}\sum_{k\sigma}\left[-\frac{1}{\pi}\mathrm{Im}G_{\sigma}(k,\omega+i\Gamma)\right]\nonumber
\end{equation}
where $\Gamma$ is the damping factor and we give an example in Fig.~\ref{fig:5}(b) where the DOS of conduction electron is the constant DOS.


\begin{thebibliography}{58}%
\bibitem{Wen} X.-G. Wen, Quantum Field Theory of Many-Body Systems (Oxford Graduate Texts, New York, 2004).
\bibitem{Kitaev1}A. Kitaev, Ann. Phys. \textbf{303}, 2 (2003).
\bibitem{Kitaev2}A. Kitaev, Ann. Phys. \textbf{321}, 2 (2006).
\bibitem{Kogut}  J. B. Kogut, Rev. Mod. Phys. \textbf{51}, 659 (1979).
\bibitem{Prosko} C. Prosko, S.-P. Lee and J. Maciejko, Phys. Rev. B \textbf{96}, 205104 (2017).
\bibitem{Nandkishore} R. Nandkishore, M. A. Metlitski, and T. Senthil, Phys. Rev. B \textbf{86}, 045128 (2012).
\bibitem{Zhong1} Y. Zhong, K. Liu, Y.-Q. Wang and H.-G. Luo, Phys. Rev. B \textbf{86}, 165134 (2012).
\bibitem{Zhong2} Y. Zhong, Y.-F. Wang and H.-G. Luo, Phys. Rev. B \textbf{88}, 045109 (2013).
\bibitem{Vijay} S. Vijay, J. Haah and L. Fu, Phys. Rev. B. \textbf{92}, 235136 (2015).
\bibitem{Parameswaran} S. A. Parameswaran, and S. Gopalakrishnan, Phys. Rev. Lett. \textbf{119}, 146601 (2017).
\bibitem{Smith} A. Smith, J. Knolle, R. Moessner and D. L. Kovrizhin, Phys. Rev. B \textbf{97}, 245137 (2018).
\bibitem{Ng} Z. Chen, X. Li, and T. K. Ng, Phys. Rev. Lett. \textbf{120}, 046401 (2018).
\bibitem{Hubbard} J. Hubbard, Proc. R. Soc. London, Ser. A \textbf{276}, 238 (1963).
\bibitem{Tsunetsugu} H. Tsunetsugu, M. Sigrist, and K. Ueda, Rev. Mod. Phys. \textbf{69}, 809 (1997).
\bibitem{Sikkema} A. E. Sikkema, W. J. L. Buyers, I. Affleck and J. Gan, Phys. Rev. B \textbf{54}, 9322 (1996).
\bibitem{Mydosh} J. A. Mydosh, P. M. Oppeneer, Rev. Mod. Phys. \textbf{83}, 1301 (2011).
\bibitem{Vojta} H. v. L\"{o}hneysen, A. Rosch, M. Vojta and P. W\"{o}lfle, Rev. Mod. Phys. \textbf{79}, 1015 (2007).
\bibitem{Coleman2015}
P. Coleman, Introduction to Many Body Physics, chapters 15 to 18 (Cambridge University Press, 2015).
\bibitem{Si} Q. Si and S. Paschen, Phys. Stat. Solid. B \textbf{250}, 425-438 (2013).
\bibitem{Coleman} P. Coleman and A. H. Nevidomskyy, J. Low Temp. Phys. \textbf{161}, 182 (2010).
\bibitem{Falicov} L. M. Falicov and J. C. Kimball, Phys. Rev. Lett. \textbf{22}, 997 (1969).
\bibitem{Assaad} F. F. Assaad, Phys. Rev. Lett. \textbf{83}, 796 (1999).
\bibitem{Czajka} M. M. Maska and K. Czajka, Phys. Rev. B \textbf{74}, 035109 (2006).
\bibitem{Stock} C. Stock et al., Phys. Rev. Lett. \textbf{121}, 037003 (2018).
\bibitem{Okazaki} R. Okazaki et al., Science \textbf{331}, 439 (2011).
\bibitem{Kotliar} G. Kotliar, S. Y. Savrasov, K. Haule, V. S. Oudovenko, O. Parcollet and C. A. Marianetti, Rev. Mod. Phys. \textbf{78}, 865 (2006).
\bibitem{Potthoff} M. Potthoff, M. Aichhorn and C. Dahnken, Phys. Rev. Lett. \textbf{91}, 206402 (2003).
\bibitem{Kennedy} T. Kennedy and E. H. Lieb, Physica A \textbf{138}, 320 (1986).
\bibitem{Kusko} C. Kusko, R. S. Markiewicz, M. Lindroos and A. Bansil, Phys. Rev. B \textbf{66}, 140513(R) (2002).
\bibitem{Zhou} Y. Zhou, K. Kanoda and T.-K. Ng, Rev. Mod. Phys. \textbf{89}, 025003 (2017).
\bibitem{Fisher} M. P. A. Fisher, P. B. Weichman, G. Grinstein and D. S. Fisher, Phys. Rev. B \textbf{40}, 546 (1989).
\bibitem{Antipov} A. E. Antipov, Y. Javanmard, P. Ribeiro and S. Kirchner, Phys. Rev. Lett. \textbf{117}, 146601 (2016).
\bibitem{Jarrell} M. Jarrell, Phys. Rev. B \textbf{51}, 7429 (1995).
\bibitem{Binder} K. Binder and D. W. Heermann, Monte Carlo Simulation
in Statistical Physics An Introduction, chapter 3 (Springer-Verlag Berlin Heidelberg, 2010).
\bibitem{Hertz} J. Hertz, Phys. Rev. B \textbf{14}, 1165 (1976).
\bibitem{Millis} A. J. Millis, Phys. Rev. B \textbf{48}, 7183 (1993).
\bibitem{Moriya} T. Moriya and A. Kawabata, J. Phys. Soc. Jpn. \textbf{34}, 69 (1973).
\bibitem{Czajka2} M. M. Maska and K. Czajka, phys. stat. sol. (b) \textbf{242}, 479 (2005).
\bibitem{Costa} N. C. Costa, J. P. Limab and R. R. dos Santosa, J. Magn. Magn. Mater. \textbf{372}, 74 (2014).
\bibitem{Peters} R. Peters and N. Kawakami, Phys. Rev. B \textbf{96}, 115158 (2017).
\bibitem{Hohenadler} M. Hohenadler and F. F. Assaad, Phys. Rev. Lett. \textbf{121}, 086601 (2018).
\bibitem{Zhong2013} Y. Zhong, Y.-F. Wang, Y.-Q. Wang and H.-G. Luo, Phys. Rev. B \textbf{87}, 035128 (2013).
\end{thebibliography}
\end{document}